\documentclass[doublecol,a4paper]{epl2}

\usepackage{amssymb,amsfonts,amsmath}
\newcommand{\eps}{\epsilon}
\newcommand{\bra}[1]{\left(#1\right)}

\title{Generation of finite wave trains in excitable media}
\shorttitle{Generation of finite wave trains in excitable media}

\author{A. Yochelis\inst{1}\footnote{Present address: Department of Chemical Engineering, Technion - Israel Institute of Technology, Haifa 32000, Israel.} \and E. Knobloch\inst{2} \and Y. Xie\inst{1} \and Z. Qu\inst{1} \and A. Garfinkel\inst{1,3}}

\shortauthor{A. Yochelis \etal}

\institute{
  \inst{1} Department of Medicine (Cardiology), University of California, Los Angeles, CA 90095, USA\\
  \inst{2} Department of Physics, University of California, Berkeley, CA 94720, USA \\
  \inst{3} Department of Physiological Science, University of California, Los Angeles, CA 90095, USA
}

\pacs{47.54.-r}{Pattern selection; pattern formation} \pacs{47.35.Fg}{Solitary waves} \pacs{47.20.Ky}{Nonlinearity, bifurcation, and
symmetry breaking}

\abstract{Spatiotemporal control of excitable media is of paramount importance in the development of new applications, ranging from biology
to physics. To this end we identify and describe a qualitative property of excitable media that enables us to generate a sequence of
traveling pulses of any desired length, using a one-time initial stimulus. The wave trains are produced by a transient pacemaker generated
by a one-time suitably tailored spatially localized finite amplitude stimulus, and belong to a family of fast pulse trains. A second
family, of slow pulse trains, is also present. The latter are created through a clumping instability of a traveling wave state (in an
excitable regime) and are inaccessible to single localized stimuli of the type we use. The results indicate that the presence of a large
multiplicity of stable, accessible, multi-pulse states is a general property of simple models of excitable media.}

\begin{document}

\maketitle

Excitable media are characterized by a large (finite amplitude) response to a supra-threshold perturbation of the rest state, followed by
decay back to the same rest state~\cite{Me:92}. Such behavior is frequently found in biological~\cite{KeSn:98}, chemical~\cite{EpPo:98},
and physical~\cite{liq_crst} systems. As suggested independently by FitzHugh and Nagumo~\cite{FH:61}, temporal excitable behavior can be
understood qualitatively via a prototype two-component ordinary differential equation of Bonhoeffer-van der Pol type with a cubic
nonlinearity in the activator field~\cite{Me:92}. In spatially extended excitable media with activator diffusion, such a perturbation
results in the formation of a solitary traveling wave, hereafter referred to as a traveling pulse~\cite{Me:92}. Since the medium returns to
the rest state after the passage of the pulse, repeated stimulation is required to generate a {\it sequence} of
pulses~\cite{EMRS:90,OGKB:00}. Here we identify a novel property of excitable systems of this type that allows us to generate economically
sequences of traveling pulses through a one-time initial stimulus, whose shape controls the number of pulses within the wavetrain.

In a comoving frame, the one-dimensional (1D) profile of a traveling pulse corresponds, in many systems, to a Shil'nikov homoclinic
orbit~\cite{Shil:65,GS:84,Ba:95,STK:98,Cham:07}. Such orbits are known to be accompanied by (an infinite number of) additional multi-pulse
homoclinic orbits~\cite{OGKB:00,GS:84,Fer:82}, and these in turn correspond to {\it spatially localized} groups of pulses (hereafter finite
pulse trains) traveling together with common speed. This speed differs in general from the speed of a single pulse. All of these states
form close to the parameter values at which the primary homoclinic orbit is present~\cite{GS:84,STK:98} and have similar speeds. In
contrast, there are also {\it spatially extended} states (hereafter infinite pulse trains) consisting of copies of the single pulse state,
in which each pulse is locked to the oscillatory tail of the preceding pulse~\cite{Ba:95,EMS:88}. Many different states consisting of
equally or unequally spaced pulses are possible, each traveling at its own speed~\cite{EMRS:90,OGKB:00,Ba:95,Fer:82,KOH:95}. Although
solutions of either type are readily constructed using ideas from spatial dynamics, their stability properties are in general
unknown~\cite{OGKB:00,Ba:95,Cham:07,Fer:82,KOH:95}. \textit{Moreover, even when the solutions are stable, no robust procedures are known
for generating a pulse train of a desired type or length without continued input.}

In this Letter, we demonstrate that in spatially extended homogeneous excitable media, distinct families of multi-pulse states can be
simultaneously stable, and explain how the different states may be generated. Of these the {\it fast} pulse trains can be generated with
great selectivity by suitably tuned one-time perturbations, via the formation of a transient pacemaker. We show how to mold the initial
perturbation to achieve the desired result. The basic idea is simple: a one-time perturbation evolves first in amplitude and so forms a
localized standing oscillation, i.e., a pacemaker. If this oscillation is in turn unstable to decay into solitary waves, the pacemaker is
transient and the number of pulses generated by the perturbation is finite. In contrast, the {\it slow} pulse trains are the result of a
long wavelength ``clumping'' instability of the traveling waves state. The results are obtained in one spatial dimension, but qualitatively
similar results hold in two dimensions.

\section{Model Equations}
In two-component models an oscillatory instability of a homogeneous state necessarily comes in with a zero wavenumber at onset and hence
produces spatially uniform oscillations~\cite{OHK:96}. This is not the case for three-component models~\cite{FOGB:00}. In the following we
focus on a system of this type that is known to support, under identical conditions, both traveling solitary pulses and spontaneous pacing
activity~\cite{Stich_thesis}:
\begin{subequations}\label{eq:FHN}
\begin{eqnarray}
\label{eq:FHNu}
      \partial_t u&=& u-u^3-v+ D\bra{\partial_{xx}+\partial_{yy}} u\,, \\
\label{eq:FHNv}
      \partial_t v&=& \eps_v \bra{u-a_vv-a_ww-a_0} \,,\\
      \partial_t w&=& \eps_w\bra{u-w}+\bra{\partial_{xx}+\partial_{yy}} w\,.
\label{eq:FHNw}
\end{eqnarray}
\end{subequations}
Here $u(x,y,t)$ can be thought of as an activator and $v(x,y,t)$ as an inhibitor; the latter is in turn controlled by a third component
$w(x,y,t)$, also generated in response to the activator. When $a_w=0$ eqs.~(\ref{eq:FHNu}) and~(\ref{eq:FHNv}) reduce to the standard
FitzHugh-Nagumo (FHN) form~\cite{Me:92}. In what follows we set $a_w=0.5$, and adopt the parameters $\eps_v=0.2$, $\eps_w=1.0$, $a_0=-0.1$,
$D=0.005$, with $a_v$ used as a control parameter.
\begin{figure}[tp]
\includegraphics[width=80mm]{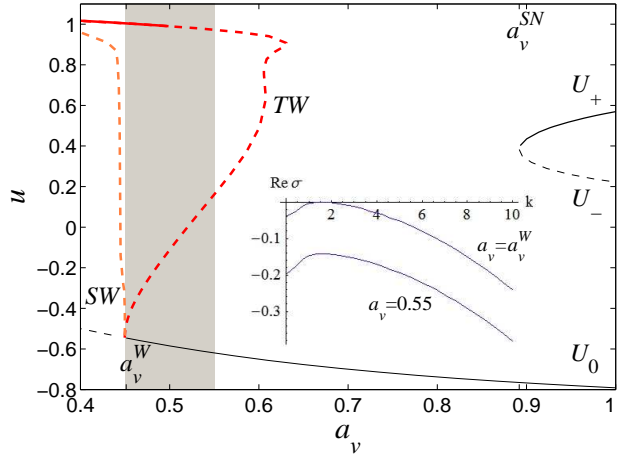}
\caption{Bifurcation diagram for spatially homogeneous steady ($U_0,U_\pm$, thin black lines) and oscillatory (heavy light colour lines)
solutions of eqs.~(\ref{eq:FHN}). Oscillatory branches of standing waves (SW) and traveling waves (TW) with fixed spatial period
$\lambda_c\equiv 2\pi/k_c\simeq 4.07$ are shown in terms of the maximum value $u_{max}$ of $u(x,t)$, and emerge from a finite wavenumber
Hopf bifurcation of $U_0\equiv(u_0,v_0,w_0)$ at $a_v\equiv a_v^W\simeq 0.4497$. Solid (dashed) lines indicate stable (unstable) solutions.
The shaded region, $0.4497\lesssim a_v \lesssim 0.55$, delimits the existence of stable pulse trains of fast type. The inset represents the
dispersion relation at onset ($a_v=a^W_v$, $k_c\simeq 1.54$, $\omega_c\simeq 0.4$, with TW phase velocity $c^{TW}_c\simeq 0.26$) and near
the rightmost limit of the shaded region, $a_v=0.55$, where $U_0$ is linearly stable.} \label{fig:bif}
\end{figure}

\section{Uniform and Nonuniform Solutions}

Equations~(\ref{eq:FHN}) admit three spatially homogeneous solutions referred to as $U_0,U_-,U_+$; the $U_\pm$ arise through a saddle-node
bifurcation at $a_v=a^{SN}_v \simeq 0.8919$ and are present for $a_v>a_v^{SN}$ only (see fig.~\ref{fig:bif}). In excitable media the rest
state, here $U_0\equiv(u_0,v_0,w_0)$, has to be stable to infinitesimal perturbations,
\begin{equation}\label{eq:lin}
\left( {\begin{array}{c}
   u  \\
   v  \\
   w  \\
\end{array}} \right) = \left( {\begin{array}{c}
   {u_0 }  \\
   {v_0 }  \\
   {w_0 }  \\
\end{array}} \right) + \delta \left( {\begin{array}{c}
   { u_k}  \\
   { v_k}  \\
   { w_k}  \\
\end{array}} \right) e^{\sigma t+ikx}+c.c.+\mathcal{O}\bra{\delta^2}.
\end{equation}
Here $\sigma$ is the (complex) perturbation growth rate, $k>0$ is the wavenumber, and $c.c.$ denotes a complex conjugate. Linear stability
analysis ($\delta\ll1$) shows that $U_0$ is stable for $a_v>a^W_v \simeq 0.4497$ (the excitable regime) but unstable for $a_v<a^W_v$ (see
inset in fig.~\ref{fig:bif}). The bifurcation at $a_v=a^W_v$ is a Hopf bifurcation with finite wavenumber $k_c\simeq 1.54$ at onset. The
corresponding critical frequency is $\omega_c\simeq 0.4$. Standard theory~\cite{CrKn:91} now shows that two branches of oscillatory states
with spatial period $\lambda_c\equiv 2\pi/k_c\simeq 4.07$ bifurcate from the rest state at $a_v=a^W_v$, a branch of traveling waves (TW)
and a branch of standing waves (SW)~\footnote{The TW were computed in the comoving frame $\xi\equiv x-c^{TW}t$, i.e., by replacing
$\partial_t$ in eqs.~(\ref{eq:FHN}) by $-c^{TW}\partial_{\xi}$, and $\partial_x$ by $\partial_{\xi}$. In this frame the TW corresponds to a
periodic orbit with period $\lambda_c$, with the speed $c=c^{TW}$ of the waves computed as a nonlinear eigenvalue using the numerical
continuation package AUTO~\cite{auto}. Temporal stability was computed via a standard numerical eigenvalue method using the time-dependent
version of eqs.~(\ref{eq:FHN}) in the comoving frame. The SW branch was computed via direct integration of eqs.~(\ref{eq:FHN}) in $0\le
x\le \lambda_c/4$ with Neumann boundary conditions at $x=0$ and Dirichlet boundary conditions at $x=\lambda_c/4$, to suppress the
instability of the SW with respect to TW perturbations.}. As shown in fig.~\ref{fig:bif}, both branches bifurcate subcritically (i.e., in
the $a_v>a_v^W$ direction) but eventually turn towards small $a_v$; for SW, the subcritical regime is very small. It follows that near
onset the TW are once unstable, while the SW are twice unstable. Moreover, the TW acquire stability with respect to wavelength $\lambda_c$
perturbations once the TW branch turns around towards smaller $a_v$, while the SW remain unstable to TW perturbations. We shall see,
however, that the TW are in fact unstable beyond the saddle-node ($a_v\simeq 0.63$) with respect to long wavelength perturbations
(fig.~\ref{fig:bif}). These instabilities accumulate at $a_v\approx 0.5$ so that the TW are stable with respect to all perturbations for
$a_v\lesssim 0.5$ only, cf.~\cite{MeAlBa:04}.

\section{Generation of Pulse Trains}

In addition to the spatially periodic SW and TW states eqs.~(\ref{eq:FHN}) also admit different types of single-pulse and multi-pulse
solutions. In a comoving reference frame these states correspond to homoclinic orbits, i.e., to solutions asymptotic to $U_0$ at $\xi \to
\pm \infty$. There are in fact two families of such states referred to in what follows as \textit{slow} and \textit{fast}, based on their
speed of propagation relative to that of the TW.

\subsection{Slow pulse trains}

\begin{figure}[tp]
(a)\includegraphics[width=80mm]{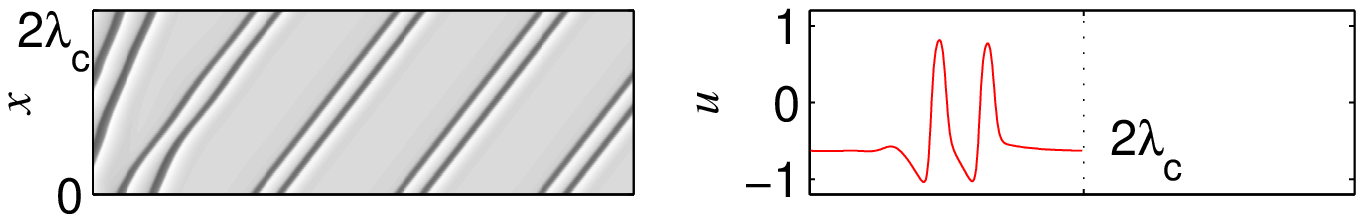} (b)\includegraphics[width=80mm]{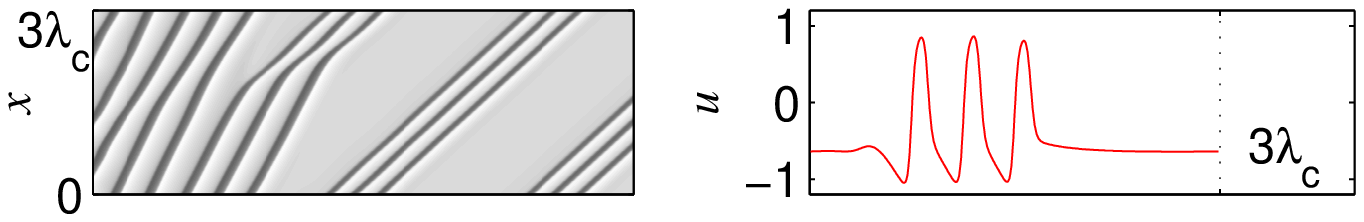} (c)\hskip 0.5mm \includegraphics[width=80mm]{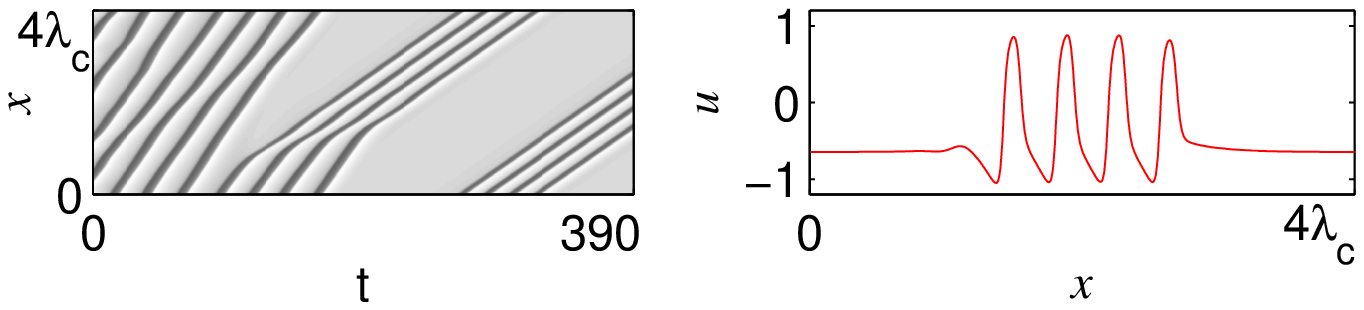}
\caption{The ``clumping'' instability of traveling waves and the formation of finite pulse trains at $a_v=0.6$ in small periodic domains
with period (a) $L=2\lambda_c\simeq 8.14$, (b) $L=3\lambda_c\simeq 12.21$, and (c) $L=4\lambda_c\simeq 16.28$. The background state filling
the domain corresponds to the homogeneous state $U_0\equiv(u_0,v_0,w_0)\simeq(-0.650,-0.375,-0.650)$. Left panels: space-time plots with
$x$ along the vertical axis and time horizontally (dark colour indicates larger values of the $u$ field). Right panels: profiles of finite
pulse trains at $t=390$, traveling without change towards the right.} \label{fig:TW_inst}
\end{figure}

Formation of the slow pulse trains is associated with the long wavelength instability of the TW (fig.~\ref{fig:bif}). These instabilities
are present in the regime $0.5\lesssim a_v\lesssim 0.63$ and the number of slow pulses that results depends on the length $L$ of the
periodic domain used to solve eqs.~(\ref{eq:FHN}). For example, for $a_v=0.6$ a bound state of two pulses is generated when the domain
period is $L=2\lambda_c$, while groups of three and four pulses form when $L=3\lambda_c$, and $L=4\lambda_c$, respectively
(fig.~\ref{fig:TW_inst}). These pulse trains propagate with speeds $c\simeq 0.08$ that are substantially slower than $c^{TW}\simeq 0.17$,
as indicated by the dramatic change in slope of the space-time trajectories of each crest as the instability evolves. The corresponding
solution profiles are shown in the right set of panels in fig.~\ref{fig:TW_inst}. No single traveling pulses are generated for this
parameter choice, presumably because the TW are stable in an $L=\lambda_c$ periodic domain.

We may think of the resulting instability as a ``clumping'' instability, and believe that it is a consequence of the fact that the fastest
growing mode in an $L=n\lambda_c$ domain is the mode with wavelength $n\lambda_c$. This property is characteristic of a phase-like
instability called the Eckhaus-Benjamin-Feir instability except that here this instability does not result in phase
slips~\cite{Janiaud:92}: instead when the wave amplitude falls locally to sub-threshold value the system nucleates the rest state $U_0$.
The result is evolution into a clumped state that {\it conserves} the number of crests.

\subsection{Fast pulse trains}

The fast pulse trains, on the other hand, are generated not by linear instability but in response to a \textit{one-time} finite amplitude
spatially localized perturbation. Such perturbations evolve more rapidly in amplitude than in phase and hence result in a spatially
localized \textit{standing} oscillation. These, like the spatially periodic SW, are in turn unstable to TW perturbations. Thus in the
shaded region in fig.~\ref{fig:bif}, suitably chosen finite amplitude perturbations evolve into transient pacemakers whose decay time
controls the length of the pulse train that is emitted. As a result the number of pulses in the wavetrain is now controlled by the decay
time of the pacemaker and hence the shape of the initial perturbation, instead of the imposed spatial period, $L$.

\begin{figure*}[htb]
(a)\includegraphics[width=101mm]{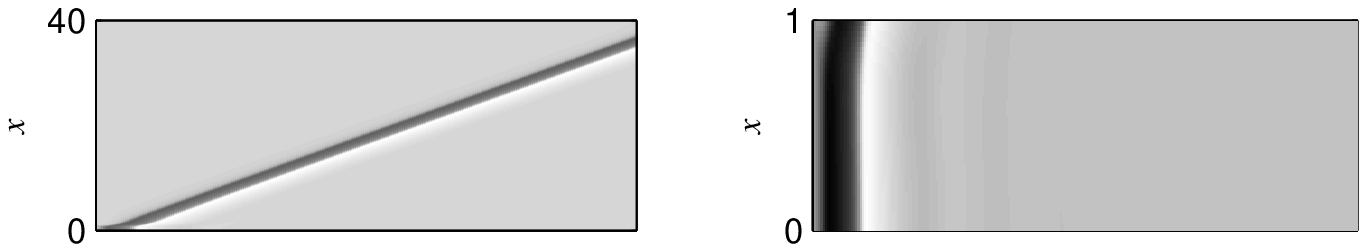} \, \, \includegraphics[width=52mm]{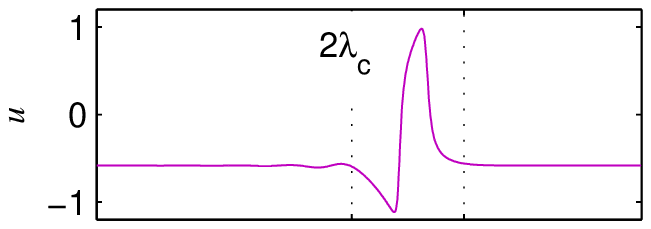}\\ (b)\includegraphics[width=101mm]{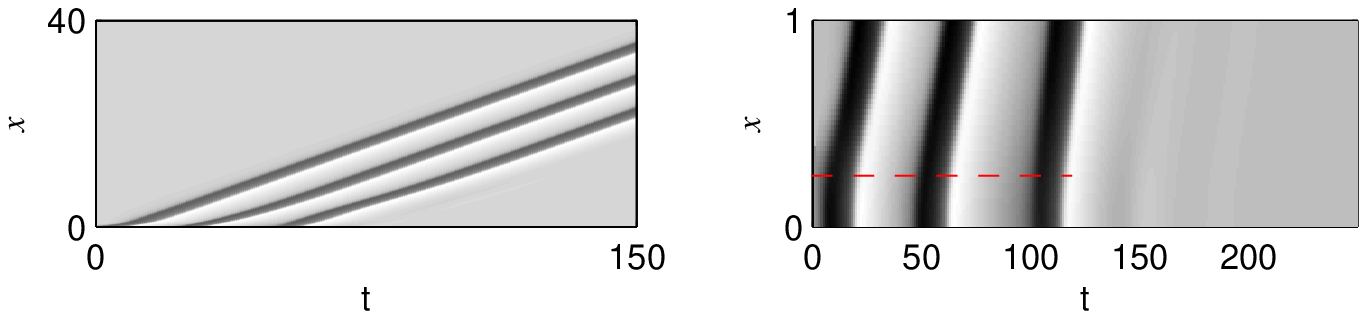} \, \,
\includegraphics[width=52mm]{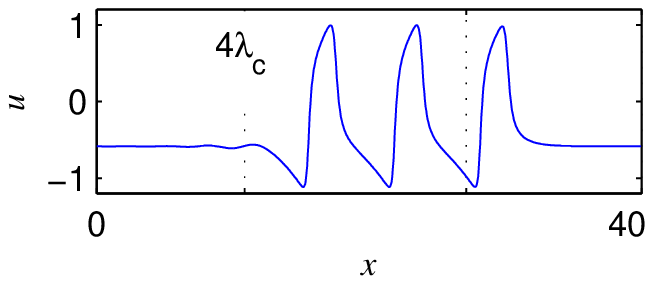}
\caption{Solutions of eqs.~(\ref{eq:FHN}) at $a_v=0.5$ with Neumann boundary conditions at $x=0$, $x=40$ using different stimuli at $t=0$:
(a) $A=0.3$, $R=1$, (b) $A=0.3$, $R=0.4$, both for $\Delta=0.1$. Left panels: space-time plots with $x$ along the vertical axis and time
horizontally (dark colour indicates larger values of the $u$ field). Middle panels: enlargement of the space-time plots in vicinity of the
initial perturbation, $x\in [0,1]$, demonstrating (a) the initial growth of the disturbance in amplitude from $u=u_0+A<0$ to $u\sim 1$,
followed by its decay when the initial perturbation is broad, and (b) the formation and eventual decay of a transient pacemaker when the
initial perturbation is narrower; the dashed line in (b) is at $x=0.25$. Each oscillation of the pacemaker generates one pulse. Right
panels: profiles of (a) single-pulse and (b) three-pulse states, both traveling without change towards the right.} \label{fig:1D}
\end{figure*}

To demonstrate the development of distinct but finite pulse trains in large domains ($L\gg\lambda_c$) by the above mechanism, we focus on
the parameter regime in which both $U_0$ and TW are linearly stable, i.e., $0.45\lesssim a_v \lesssim 0.5$. In what follows, we consider
eqs.~(\ref{eq:FHN}) with Neumann boundary conditions on the 1D domain $0\leq x\leq L$, and examine the effect of different initial
perturbations on pacemaker emergence and the resulting solutions. The perturbations have a smoothed top-hat profile centered at $x=0$,
\begin{equation}\label{eq:pert}
     u\bra{x,t=0}=u_0+\frac{A}{2}\bra{1-\tanh\frac{x-R}{\Delta}}\,,
\end{equation}
where $A$ denotes the amplitude of the perturbation, $R\ll L$ its width, and $\Delta\lesssim R$ measures the steepness of the initial interface. When
$a_v>a_v^W$ and $A\sim 0.1, R\sim 1$, the excitability threshold is locally exceeded and the system responds with a single propagating
pulse. Figure~\ref{fig:1D}(a), shows the propagation of such a pulse in a space-time plot (left panel) with an enlargement of the initial
phase (middle panel) and the spatial pulse profile (right panel). The figure shows that the initial perturbation grows abruptly in place
before decaying into a propagating pulse. Once in motion, the profile of the pulse becomes distinctly asymmetrical, with a train of
spatially decaying oscillations towards its rear, as expected of a Shil'nikov saddle-focus~\cite{Shil:65}.

While such single pulse profiles are known from the two-variable FHN system~\cite{Me:92,Cham:07}, the three-variable model
[eqs.~(\ref{eq:FHN})] allows the generation of more complicated solutions such as $N$-homoclinic orbits of Shil'nikov type~\cite{GS:84}. To
observe these states we decrease the stimulus width to $R=0.4$. This initial perturbation sets up a transient pacemaker and the system now
generates a finite pulse train [fig.~\ref{fig:1D}(b), left panel], even though no additional stimulation is applied after $t=0$. The middle
panel of fig.~\ref{fig:1D}(b) shows that the pacemaker oscillates in the form of a standing oscillation in $0\le x\lesssim0.25$ for three
cycles, before decaying. Each oscillation cycle that exceeds the excitability threshold results in the generation of a traveling pulse.
Consequently an initial stimulus with $A=0.3$, $R=0.4$ triggers three pulses moving together as a group [see fig.~\ref{fig:1D}(b)].

Numerically, we find that the single pulse in fig.~\ref{fig:1D}(a) propagates with speed $c^{pulse}\simeq 0.2482$, while the pulse
\textit{trains} all propagate with essentially identical and slightly larger speed $c\simeq 0.2487$, to within numerical error of order of
$10^{-5}$. Both velocities are substantially larger than the phase velocity of the TW, $c^{TW}\simeq 0.1877$ at $a_v=0.5$. Consequently we
call the resulting finite pulse trains {\it fast}. Comparison with fig.~\ref{fig:TW_inst} shows that the fast pulses are much broader than
the slow pulses, but otherwise resemble each other.

\begin{figure}[htb]
\includegraphics[width=70mm]{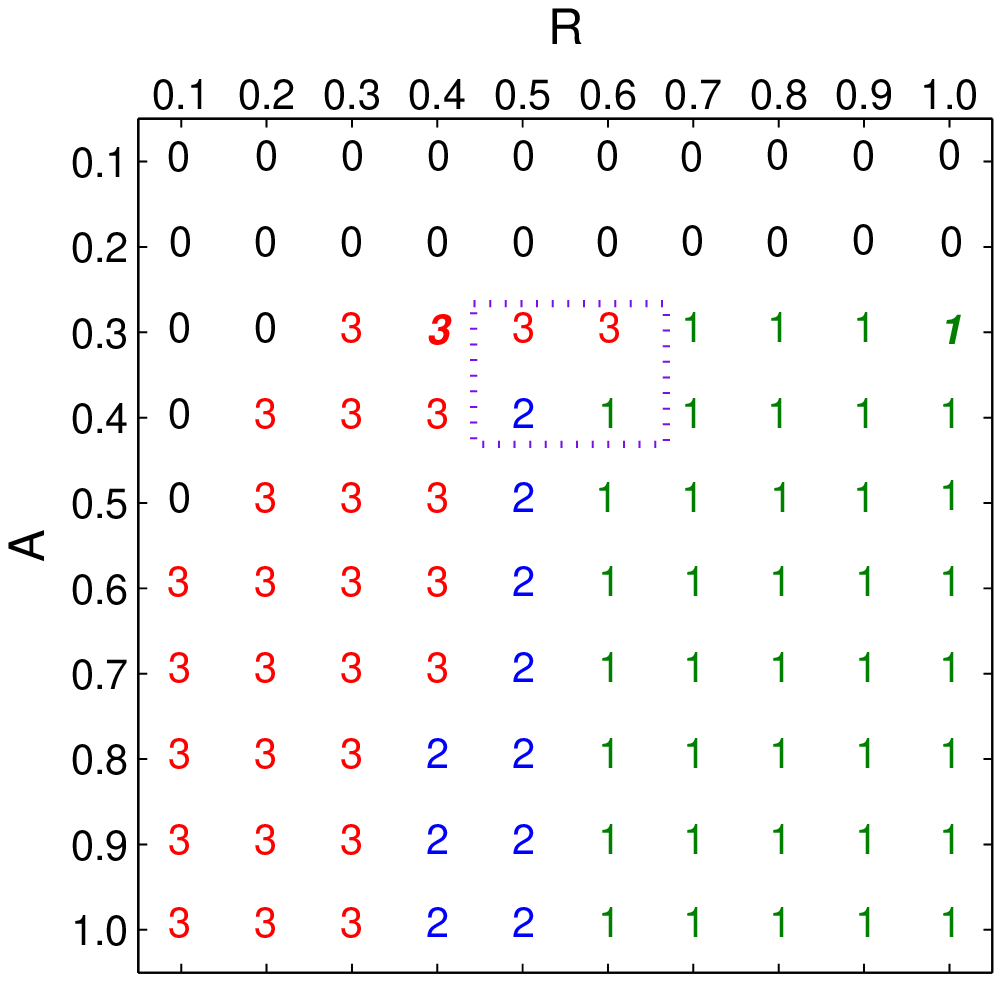}\\
\includegraphics[width=80mm]{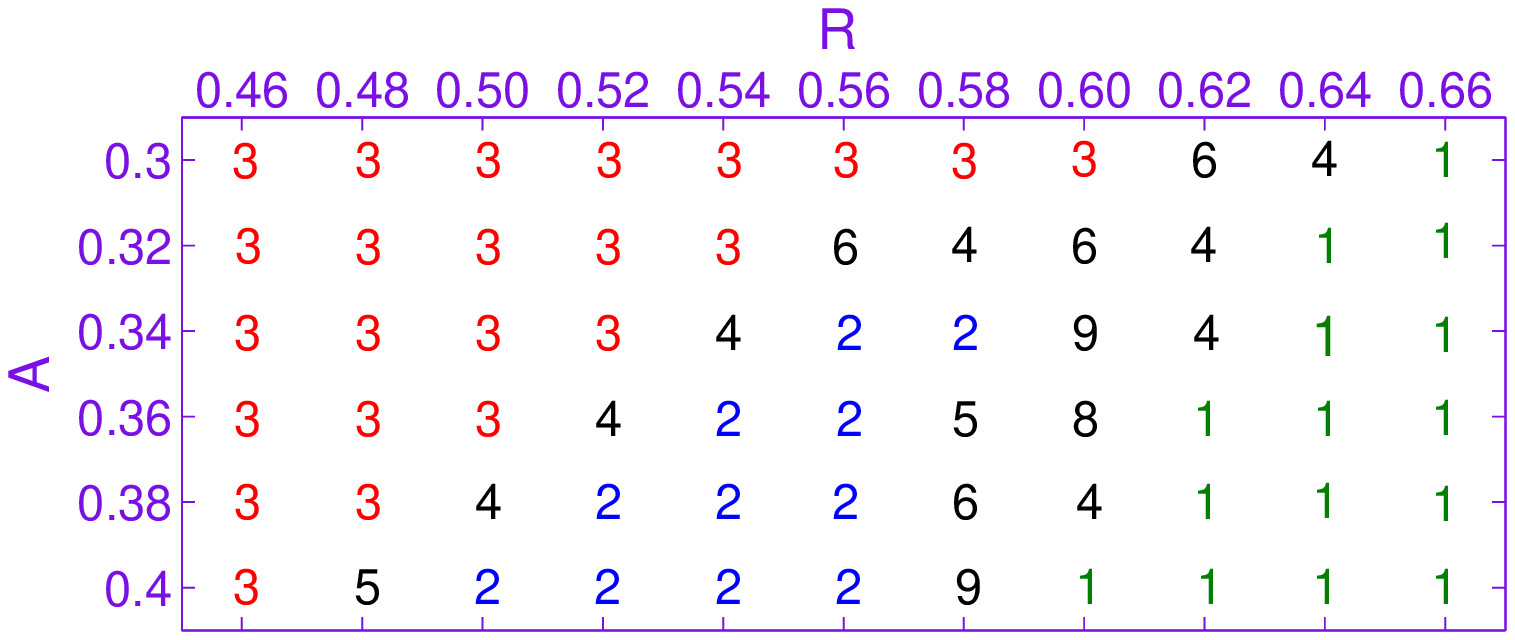}
\caption{Top panel: finite fast pulse trains computed from the initial perturbation~(\ref{eq:pert}) with $\Delta=0.1$ for different values
of $A$ and $R$ when $a_v=0.5$. The integers correspond to the number $N$ of pulses in the wavetrain, the bold italic fonts corresponding to
the solutions described in fig.~\ref{fig:1D}. Bottom panel: variation in $N$ in the vicinity of the extinction boundary, in the region
indicated by the small rectangle in the top panel. The numerical integration of eqs.~(\ref{eq:FHN}) was performed with $\Delta x=0.005$ and
Neumann boundary conditions on an $L=50$ domain.}\label{fig:A_R}
\end{figure}

The finite pulse trains accessible by varying the amplitude $A$ and width $R$ of the initial perturbation are summarized in
fig.~\ref{fig:A_R}. In the vicinity of the extinction boundary the number of pulses increases, albeit in a nonmonotonic fashion, and this
property of the system can be exploited to identify empirically the initial conditions that result in the desired number of pulses. In the
bottom panel of fig.~\ref{fig:A_R} we provide a more detailed sample of the possible outcomes. Evidently the number of pulses is very
sensitive to the form of the initial perturbation, and this sensitivity extends to the dependence of the results on the length scale
$\Delta$ introduced in eq.~(\ref{eq:pert}).

Qualitatively similar results are found for other parameterized families of initial conditions, provided these broadly resemble the
top-hat profile in eq.~(\ref{eq:pert}).

\section{Bistability between Different Pulse Trains}

\begin{figure}[tp]
\includegraphics[width=80mm]{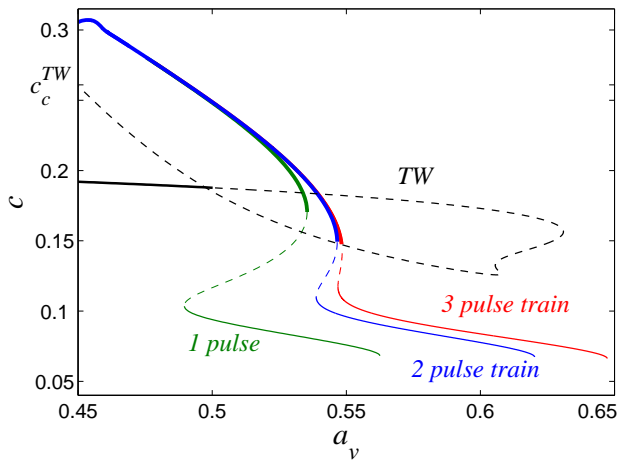}
\caption{The speed $c$ of the fast (heavy colour lines) and slow (thin colour lines) pulse trains as a function of $a_v$ together with the
phase speed of traveling wave (TW) states. Only three branches of traveling pulse trains are shown. The phase speed of the TW lies between
the speeds of the fast and slow pulse trains. Solid (dashed) lines indicate stable (unstable) states. The accumulation of the saddle-nodes
of the fast pulse trains ($a_v\approx0.55$) defines the upper boundary of the shaded region displayed in fig.~\ref{fig:bif}. The pulse
trains were computed in a comoving frame, on a periodic domain with $L=100$.}
\label{fig:bif_hom}
\end{figure}

The similarity in the pulse profiles suggests that the two families of finite pulse trains may in fact be related. This supposition is
supported by our numerical simulations, which show that stable pulse trains of slow and fast type may coexist, cf.~\cite{Stich_thesis}. To
identify the connection between these two families we have again employed numerical continuation~\cite{auto} but this time initiated with
single, double and triple homoclinic profiles of fast type, such as those shown in fig.~\ref{fig:1D}. In fig.~\ref{fig:bif_hom}, we show
that the slow and fast pulse trains are in fact separated by unstable branches (dashed lines). This result explains why we find two
families of traveling pulse trains, and why we observe bistability between them. Observe that the fast pulse trains all travel with
essentially the same speed $c$ as $a_v$ decreases, and do so over a broad interval of $a_v$ values. This is the case for all the
multi-pulse states shown in fig.~\ref{fig:A_R}. This fact facilitates the insertion of additional pulses into a pulse train through
incremental changes in the initial stimulus. The resulting selectivity is a consequence of the large multiplicity of coexisting stable
pulse trains and the intertwined structure of their basins of attraction. In contrast, the branches of slow pulse trains all turn around
towards smaller $a_v$ as $c$ decreases (not shown), and in so doing follow the behavior of the TW branch.

The results in fig.~\ref{fig:bif_hom} help us to understand the observed absence of fast pulse trains at $a_v=0.6$, and suggest that the
empirically determined upper boundary of the shaded region in fig.~\ref{fig:bif}, i.e., the region within which finite amplitude
perturbations generate finite pulse trains, is determined by the accumulation point of the saddle-node bifurcations at which the {\it fast}
pulse trains lose stability and cease to exist. As seen in the figure these saddle-node bifurcations accumulate at $a_v\simeq 0.55$, an
observation that can be confirmed by numerical continuation of the additional multi-pulse states shown in fig.~\ref{fig:A_R}. This value is
close to that at which the TW solutions acquire stability with respect to wavelength $n\lambda_c$ ($n>1$) perturbations but is not related
to it. Numerically we find that in $a_v \lesssim 0.55$ the finite amplitude perturbations always reach fast pulse trains, even when these
coexist with stable slow pulse trains. Figure~\ref{fig:bif_hom} also shows that for $a_v\gtrsim 0.55$ slow pulse trains, including the
missing slow single-pulse state in $L=\lambda_c$ domains, should be reachable by appropriate finite amplitude perturbations of $U_0$.

\section{Generation of Finite Trains of Target Waves}

\begin{figure}[tp]
\includegraphics[width=80mm]{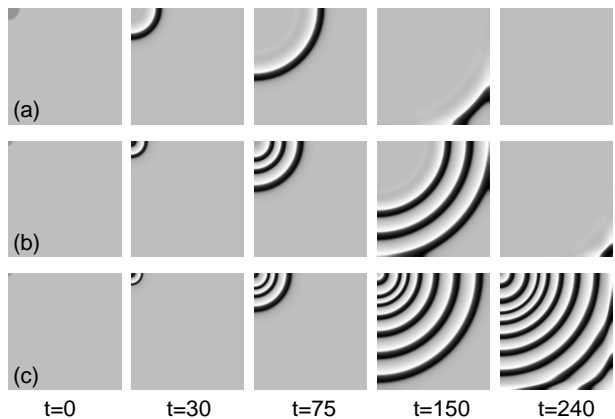}
\caption{Solutions of eqs.~(\ref{eq:FHN}) in the $(x,y)$ plane with $a_v=0.5$, using different initial stimuli. Dark colour implies larger
values the $u$ field; the propagation direction is from the top-left corner to the bottom-right corner. (a) Single wave propagation for
$R=2.0$. (b) Transient pacing and the formation of a finite number of target waves for $R=0.8$. (c) Formation of a stable pacemaker and
persistent target wave emission for $R=0.6$. In all cases the perturbation amplitude is $A=0.3$, with $\Delta=0.1$; the physical domain
size is $20\times20$ with Neumann boundary conditions, and $\Delta x=\Delta y=0.005$.}
\label{fig:2D}
\end{figure}

We now turn to the corresponding results in two spatial dimensions (2D), with Neumann boundary conditions on a square domain. The initial
perturbation is taken to be spot-like and centered at the top-left corner of the domain, as shown in fig.~\ref{fig:2D}; the parameter $R$
now measures its extent in the radial direction.

The two types of behavior, propagation of a single target wave [see fig.~\ref{fig:2D}(a)] and of a finite train of target waves due to
transient pacing activity [see fig.~\ref{fig:2D}(b)], persist also in 2D; the spatial cross-sections of these states along the propagation
diagonal [figs.~\ref{fig:2D}(a,b)] admit profiles that resemble the profiles of fast family type obtained in 1D (see fig.~\ref{fig:1D}).
The value of $R$ required for the finite target waves is now larger ($R \sim 1$) than in the 1D case, but with increasing $R$ single-pulse
propagation results, as in 1D. However, in contrast to 1D a one-time perturbation with a small value of $R$ produces {\it persistent}
pacing activity that results in standard target waves [see fig.~\ref{fig:2D}(c)]~\cite{Stich_thesis}. These results are not affected by the
choice of partial domain and apply equally to a large square domain with a spot-like perturbation in the center. We attribute the
differences between the 1D and 2D results to curvature effects in the pacemaker region.

\section{Conclusions}

We have demonstrated that stable multi-pulse solutions can be generated in a spatially extended excitable medium by a \textit{one-time}
stimulus, with no need for additional stimulation. These finite pulse trains are the result of the formation of a transient pacemaker with
decaying amplitude that generates solitary waves while its amplitude remains above threshold for pulse generation. The primary requirement
for this property of the system is a nonlinear (subcritical) instability of the uniform state to both traveling waves (TW) and standing
waves (SW), as in fig.~\ref{fig:bif}. We have seen that under these conditions stable multi-pulse states are present, and that the basins
of attraction of these states can be selected by a suitably tailored initial stimulus through the excitation of a transient pacemaker. We
believe these results to be independent of the details of the system used here, and thus characteristic of a broad class of excitable
systems.

In the present example the pulse trains generated by the transient pacemaker propagate with a speed substantially larger than the
coexisting spatially periodic TW, and may also coexist with a distinct family of traveling pulses that propagate more slowly than the TW.
The latter form as a result of a long wavelength ``clumping'' instability of the TW that is also a consequence of the subcriticality of the
TW branch. However, the slow pulse trains are not accessible by finite amplitude perturbations of the rest state of the type we use.

The study was conducted within the phenomenological framework of a three-variable FitzHugh-Nagumo type model~\cite{Stich_thesis}, and
employed bifurcation analysis, numerical continuation, and parametrized stimuli to generate a variety of finite pulse trains, as summarized
in fig.~\ref{fig:A_R}. These results demonstrate that near the extinction boundary the length $N$ of the fast pulse train becomes a
sensitive function of the shape of the initial stimulus, enabling us to ``dial in'' different pulse trains by appropriately shaping the
initial stimulus. The details of the resulting look-up table (fig.~\ref{fig:A_R}) are system-specific but need only be computed once;
similar look-up tables can be established empirically for each model of interest. As shown in figs.~\ref{fig:2D}(a,b) these results extend
to two space dimensions as well.

We speculate that this intriguing property of excitable systems may provide new insights into a number of applied science processes
exhibiting reaction-diffusion behavior~\cite{Me:92,FOGB:00,MGS:06}, including chemical reactions~\cite{EpPo:98}, excitation and propagation
of signals in neurons and cardiac tissue, and intracellular calcium cycling~\cite{KeSn:98}. Technological applications to information
transfer using nonlinear optical systems~\cite{CDT:98} and to unconventional computation based on reaction-diffusion systems~\cite{IGG:06}
are also possible.

\acknowledgments This work was supported by NIH/NHLBI Grant P01 HL078931 and by National Science Foundation Grant DMS-0605238.

\end{document}